\documentclass[a4paper,twoside]{article}
\newcommand{\affil}[1]{$^{\rm #1}$}
\usepackage[authoryear]{natbib}
\bibpunct{(}{)}{;}{a}{}{,}
\usepackage{graphicx}
\date{} 
\title{\large\bf\flushleft NGC 5128: The Giant Beneath  }
\author{\parbox{\textwidth}{\flushleft
\vspace{-0.5cm}
{\it Gretchen L. H. Harris\affil{A,B}}\\
\vspace{0.4cm}
{\small \affil{A}\,Department of Physics and Astronomy, University of Waterloo, Waterloo ON N2L 3G1, Canada}\\
{\small \affil{B}\,Email: glharris@astro.uwaterloo.ca}}}
\begin{document}
\twocolumn[
\begin{changemargin}{.8cm}{.5cm}
\begin{minipage}{.9\textwidth}
\vspace{-1cm}
\maketitle
%
%
\small{\bf Abstract:}

I review what we have learned about the old stellar population
of NGC 5128, the only large E galaxy close enough that we can currently
observe individual stars as faint as the horizontal branch.  Although
its galaxy type is still a matter of debate for some, the uncertainties
over distance are now largely resolved; comparison of five stellar
distance indicators gives d=3.8$\pm$0.1 Mpc.  The globular
cluster system, which was once perplexingly invisible, is now known to be 
predominantly old with a substantial metal-rich component. 
The GCS total population and luminosity function are normal.
and the clusters follow the same fundamental plane relation as
those in the Milky Way and M31.   Finally, the halo
out to at least $\sim 7r_{eff}$
is dominated by metal-rich stars which are also predominantly old, 
with age and metallicity tantalizingly similar
to the majority of globular clusters.

\medskip{\bf Keywords:} Galaxies: distances --- galaxies: stellar 
content and star clusters ---
galaxies: individual (NGC 5128)

\medskip
\medskip
\end{minipage}
\end{changemargin}
]
\small
\section{Introduction}
In this review I will discuss early problems
which confused our understanding of NGC 5128 
and provide an overview of what we have learned about
the properties of its old stars, with an emphasis 
on the globular cluster and field star populations.
Because NGC 5128 is $<$ 4Mpc away ($(m=M)_0 = 27.9$) we can study its 
stellar component in greater depth and detail than is possible
for any other large elliptical galaxy.  Consequently we have data for 
hundreds of globular clusters and planetary nebulae
which are telling a rich history of how and when its stars formed.
In addition (and unknown to many readers) 
we can now resolve individual halo stars, long
period variables and Cepheids in NGC 5128, telling us its distance
 and providing additional clues to its history. 
For recent results on these and other stellar components see:
distance \citep{har10}, ages of field stars and long period
variables \citep{rej03, rej05}, 
globular clusters (\citep{peng04b, woodoz, wood10},  
and planetary nebulae \citep{wal99, peng04a, wal10}.

\section{The Nearby E Galaxy}
When most people look at NGC 5128 they see a peculiar 
galaxy dominated by a prominent 
dust lane  and ignore the rest.  What I see is a nearby E galaxy 
whose oldest stars can tell us a great deal about its history.  
A quick look at the literature over the past 60 years shows a growing 
body of research mostly on the radio jets, dust lane, Xray sources
and much less on the dominant baryonic component which is
the E galaxy beneath.  But, particularly in the past twenty years, 
we are now beginning to understand 
the {\it stellar} history of NGC 5128 as well.

\subsection{Galaxy Type} 
Because NGC 5128 is associated with a strong radio source and its physical
appearance in the center is dominated by a wide absorption band, 
this galaxy continues to
be perceived as one whose fundamental properties are likely
to be atypical.  
But as early as the mid-1950s studies indicated that, 
aside from the dust lane, the
global, structural properties 
of NGC 5128 were consistent with those 
of an E galaxy. \citet{baade54} called it ``an unresolved E0 nebula
with an unusually strong and wide central absorption band'' and concluded
that ``the two nebulae...form a close pair in a state of strong 
gravitational interaction, perhaps actually in collision''.  
Based on data from 
photographic plates in three wavelength bands, \citet{ser58} found
isophotal countours which were remarkably similar in all colours
and fitted a deVaucouleurs profile out to 10$^{\prime}$ ($\sim$ 10kpc)
from the centre of the galaxy.   

In their review \citet{ebneter83} argue that ``Cen A has a probably 
undeserved reputation for being one of the most peculiar galaxies
in the sky'' and point out that many of its characteristics would 
not be as visible or obvious if the system were more distant.  
And in his study of both the halo light distribution and
 the dynamics of the dust lane region,
 \citet{gra79} concludes the ``the main body of NGC 5128 resembles in
many respects a normal giant elliptical galaxy'' and ``the main
radio characteristics of NGC 5128 and the energetic phenomena...are
a consequence rather than the cause of the unusual structural features 
of the galaxy as a whole''. 

In spite of this, there is still no clear consensus on 
what type
to use for 
NGC 5128 (cf. Table 1); the ``vote'' 
is clearly split between S0p and E0p, presumably because
of the dust lane.  \citet{mor58} suggested that the S0 classification is
applied to galaxies which are often very different from each other and 
\citet{vdb90} agrees that ``the S0 classification type comprises a number 
of physically quite distinct types of objects that exhibit only
superficial morphological similarities''.  More recently 
he indicates that ``NGC 5128 is an object that does not find a natural
home in the Hubble classification scheme'' (\citet{vdb09},
 private communication). In this discussion I 
will consider NGC 5128 an Ep.

\begin{table}[h]
\begin{center}
\caption{Galaxy Type for NGC 5128}\label{table}
\begin{tabular}{lc}
\hline galaxy type & source \\
\hline
\\
like NGC 3379 (E1) & \citet{ser58} \\
E0p?  &  \citet{mor58} \\
(E0~+~Sb?) & \citet{san61} \\
S0p             &  \citet{fre70} \\
E0p      &  \citet{vdb76} \\
E~+~spiral   &  \citet{duf79} \\
S0    &  \citet{santam81} \\
S0p   & NED database (2009) \\
Ep   &  this paper \\
\hline
\end{tabular}
\medskip\\
\end{center}
\end{table}

\subsection{Distance}
It may seem a bit odd that several decades passed before 
we could determine a reliable distance to a galaxy not far beyond the 
boundaries of the Local Group.  But NGC 5128 was (until recently)
too  distant to apply most 
stellar candles, and is too close to use the Hubble Law with confidence
or to observe as we do other large E galaxies.  But in
the past two decades we have been able to 
 derive a distance based on the properties of its stellar component.  

As can be seen in Figure 1,  
distance estimates before $\sim$1980 
ranged from 2 to 9 Mpc, a factor of $>$4! And this 
apparent dichotomy between large and small distances often led to
the use of an ``average'' value of $\sim$5 Mpc. 
But since the late 1980s it has been possible to use a variety of
 stellar distance indicators: luminosity of the red giant branch tip (TRGB),
planetary luminosity function (PNLF), long period variables (LPV), surface
brightness fluctuations (SBF), and Cepheid variables.  The quality of the data 
and the calibration of these methods now support a 
distance of 3.8$\pm$0.1 Mpc for NGC 5128, and a full review
is given in this volume \citep{har10}. 

\begin{figure}[h]
\begin{center}
\includegraphics[scale=0.30, angle=0]{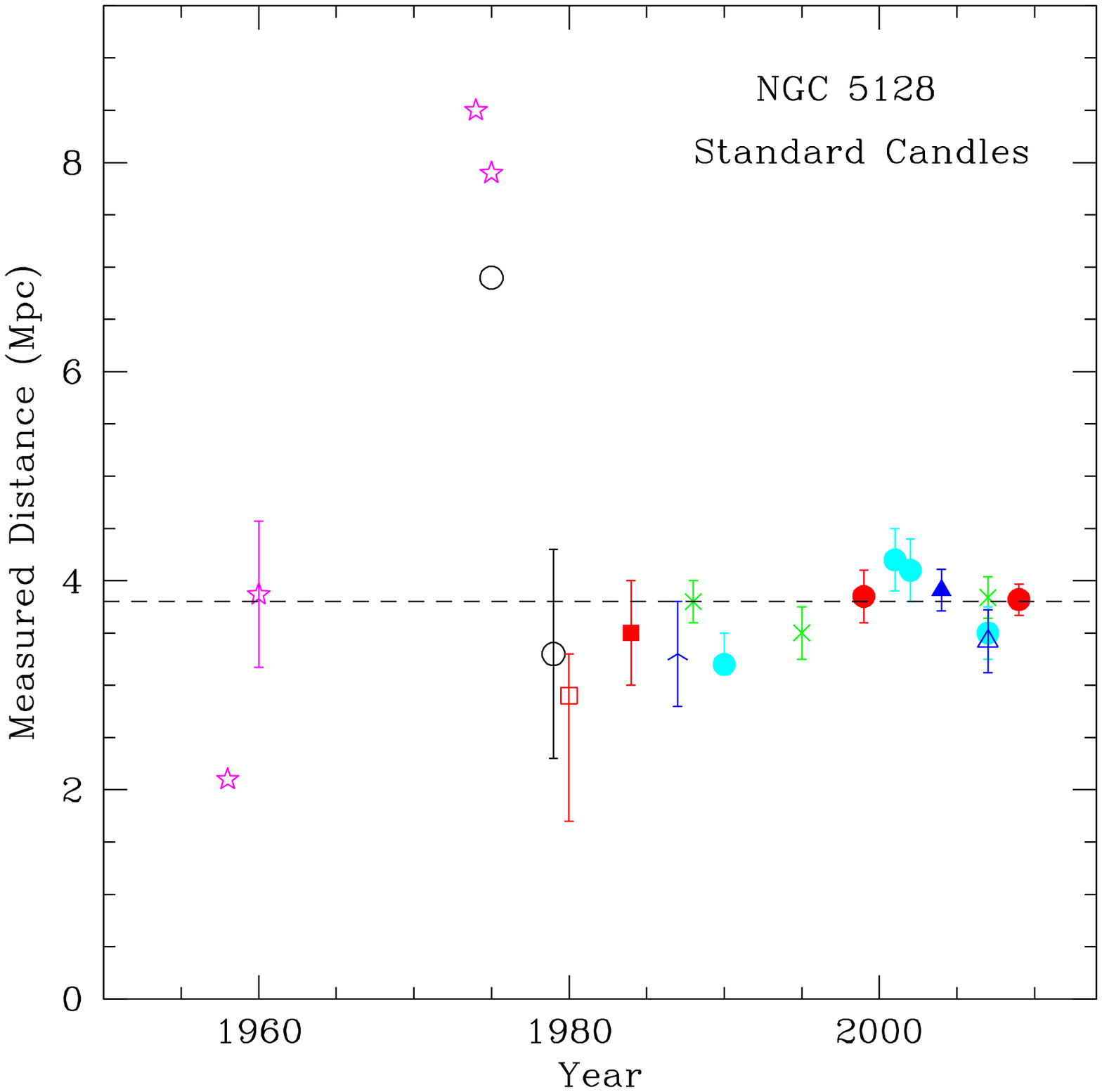}
\caption{Distance to NGC 5128 vs. year of publication:
H II regions (magenta stars), redshift (open black circles), globular clusters
(red squares), SN1986g (blue chevron), TRGB (red circles), 
SBF (cyan circles), PNLF 
(green stars),
LPV (filled blue triangle), and Cepheid variables 
(empty blue triangle).  Error bars are plotted when available.
For sources see \citet{israel98}, \citet{rej04}, \citet{har10} and
references  
given therein.  The horizontal line is the current best value of d=3.8$\pm$0.1
\citep{har10}.} 
\label{fig.1}
\end{center}
\end{figure}

\section{The Globular Clusters}

An important factor in the confusion about the nature of NGC 5128 was 
its apparent lack of globular clusters.  By the late 1970s studies of  
galaxies beyond the Local Group had found  
globular clusters (GCs) in every galaxy searched, and shown
that large E galaxies contained thousands of GCs (cf. \citet{har79}
and references cited therein).  But in the case
of NGC 5128 no such GC population had been uncovered. 

\begin{figure}[h]
\begin{center}
\includegraphics[scale=0.30, angle=0]{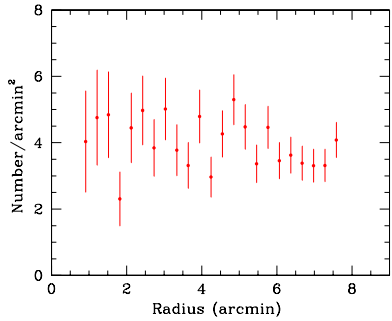}
\caption{Star counts in NGC 5128 based on data from \citet{vdb79}.
 }\label{fig.2}
\end{center}
\end{figure}

\citet{vdb79} carried out star counts in annuli between 45$^{\prime \prime}$
($\sim$ 800pc) and 484$^{\prime \prime}$ ($\sim$9000pc) 
(cf. Figure 2), finding an excess of only 15$\pm$60 objects above background.
An obvious explanation for the apparent lack of GCs in NGC 5128 was the
dust lane and its signalling of a cataclysmic history.  We now know that the
 GCs are present in numbers expected for an E galaxy in a small group 
(e.g. \citet{har84}, \citet{har04b}).
But they are hard to separate unambiguously from the noise
of foreground Milky Way halo stars and compact background galaxies.
The two main reasons for this are proximity and low Galactic
latitude.  At a distance of 3.8 Mpc, 1$^{\prime}$ corresponds to  
$\sim$1100pc and, consequently, a survey to discover a widely distributed
population like globular clusters must cover an area of more than two 
square degrees.  Observationally this is a major challenge, but that 
is not the worst.  

Since globular clusters in galaxies well beyond the Local Group generally 
appear starlike, they cannot be distinguished from either foreground
field stars or background galaxies by appearance alone.  
To reduce contamination, we start by using the fact that
clusters occupy a well defined range in
colour and luminosity to detect them photometrically
and eliminate a large fraction of non-cluster sources. 

But, for NGC 5128 this is nearly useless.  
Its brightest GCs should 
appear at an apparent magnitude of $V \sim$17 while  
the peak of the GC luminosity function (GCLF)
would be seen at a magnitude of $V \sim$21.  
Because of NGC 5128's low Galactic latitude (19$^o$) we see it through
the Milky Way halo and disk and its mainly late type stars whose colours
{\it and} apparent magnitudes  
will fall in the same range as GC candidates.  Compact background galaxies
which are fairly easily seen through the dilute halo may also have GC-like 
colours and surveys show that their numbers begin to increase rapidly 
at $V \sim 20-21$, just at the GCLF peak \citep{met01}. 
In short, the sample contamination problem is extraordinarily 
difficult to solve.

The GC mystery began to disappear with the discovery by \citet{gra80} of 
a globular cluster candidate $\sim 8.2^{\prime}$ northeast of
the galaxy centre.  This was followed quickly with the identification of
six more \citep{vdb81}.  All seven were initially selected by their slightly
non-stellar appearance and then confirmed spectroscopically as having 
radial velocities consistent with membership in NGC 5128. 
Using Schmidt plates with a much larger field of view than
the  
4-m prime focus plate used by \citet{vdb79}, \citet{vdb81}
also revisited the question of star counts.
Their results, reproduced in Figure 3,
 show a clear excess of objects within $\sim 375
^{\prime \prime}$ of the nucleus; based on this they estimated the total cluster
population at 600$\pm$61.
Later wide field surveys have estimated the total GCS population 
at $\sim$1000-2000 \citep{har84,har04b}

\begin{figure}[h]
\begin{center}
\includegraphics[scale=0.30, angle=0]{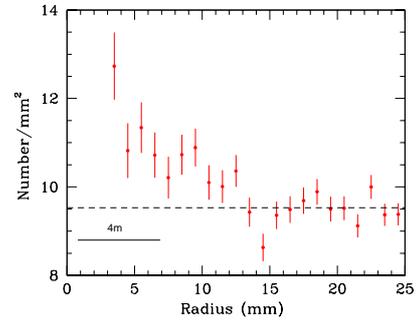}
\caption{Surface density of objects plotted vs.
distance from the centre of NGC 5128, 
adapted from \citet{vdb79} (Figure 2); note
the line labeled 4m indicating the area counted 
previously by \citet{vdb79}.}\label{fig.3} 
\end{center}
\end{figure}

\subsection{GCS Total Population}
\begin{figure}[h]
\begin{center}
\includegraphics[scale=0.30, angle=0]{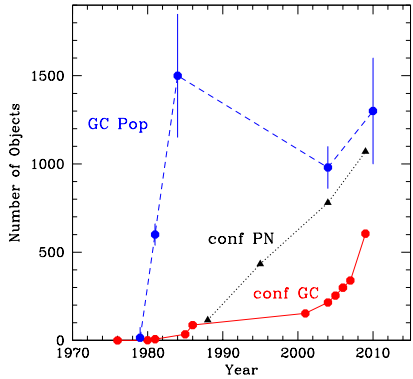}
\caption{Plotted are estimates of total GC population (blue), GCs confirmed
by radial velocity and/or resolved by HST imaging, and confirmed planetary
nebulae.  If the estimated  total number of GCs in NGC 5128 is
correct (cf. Figure 5 and accompanying text),
 then we have now identified $\sim$ half of the GC population.}  
\label{fig.4} 
\end{center}
\end{figure}

Once the existence of globular clusters in NGC 5128 was established, 
increased efforts were made to understand this system.  But the 
observational issues described above have made it extremely difficult to
build an overall picture.  To illustrate this I have plotted a historical
overview in Figure 4.  Shown plotted vs. year  
are: estimates of the total GC population
(\citet{vdb79,vdb81,har84,har04b} and this paper); 
number of confirmed 
GCs \citep{gra80,vdb81,hes86, 
rej01,peng04b,
wood05,wood07,bea08, 
woodoz}; and, for comparison,
the number of confirmed planetary nebulae \citep{jac88,ford89, 
hui93,peng04a,wal10}. 
Note that the number of known GCs reaches a plateau at $\sim$100 in the late 
1980s and remains virtually unchanged for more than a decade.  This reflects the
fact that the ``easy'' GC candidates which could be identified by a slightly
non-stellar appearance had pretty much been found.  And remember,
adding more to the sample by photometry alone is an almost impossible task 
for ground based seeing, while the large angular scale of the system
precludes HST imaging due to its small field of view.
In their wide field imaging study 
\citet{har04a} found $\sim10^5$ objects which
fell in the colour and magnitude range populated by globular clusters in 
NGC 5128.  If the total cluster population is $\sim$2000, it is embedded in
a S/N of 0.02; i.e. only about 2$\%$ of the detected star-like objects
in the field encompassed by the halo of NGC 5128 are likely globular
clusters.  Recent increases in the number of known clusters have been 
achieved by the brute force approach of obtaining radial velocity
spectra for the brightest candidates \citep{bea08, wood05, wood07,wood10}
and the number of confirmed GCs is now 607.
\citet{wood10} fit a Gaussian to the luminosity function (LF) of
their sample, finding a turnover of $T_1 (to)$ = 19.44 mag.
This is almost 1 mag brighter than the expected value and a clear sign that 
this sample is missing most of the fainter clusters.  Because both the
distance to NGC 5128 and the absolute magnitude of the standard globular
cluster luminosity function (GCLF) turnover are well known we can force
a fit to the cluster sample, assuming it is complete for the brightest 
clusters, and estimate the total population of the cluster system, $N_t$. 
With $(m-M)_0 = 27.9$ \citep{har10}, $A_V = 0.35$ \citep{schlegel98},
and $M_V (to) = -7.3$ for giant ellipticals \citep{har01} the GCLF 
turnover should occur at $V = 20.95$ or $T_1 = 20.47$.  Fitting a Gaussian
function with $\sigma = 1.4$, typical of gE's \citep{har01}  yields the result 
shown in Figure 5, from which we estimate 
$N_t = 1300 \pm 300$.

This value is consistent with previous estimates, but is based on a more
 secure data set, especially for the brightest clusters, and stronger
constraints on the distance to NGC 5128.  From Figure 5 it is also clear
that roughly half the GC population is now known, and that almost all of 
the undiscovered clusters will be fainter than $V \sim 20$ or $T_1 \sim 19.5$.
Finally, we can calculate the specific frequency 
$S_N = N_t 10^{0.4(M_V ^T +15)}$ \citep{har81}.  Combining 
$N_t = 1300 \pm 300$ and
$M_V ^T = -22.1$ \citep{har04b}, gives $S_N = 1.9 \pm 0.5$. 
Disk galaxies typically have low $S_N \leq 1.5$, E galaxies have $S_N$ ranging from 1-10 and even higher, while field E galaxies are on the lower end of
this range with $S_N \leq 2$ \citep{har01}.  Thus $S_N \sim 2$ for NGC 5128
is consistent with a normal GCS in a small group.

\begin{figure}[h]
\begin{center}
\includegraphics[scale=0.38, angle=0]{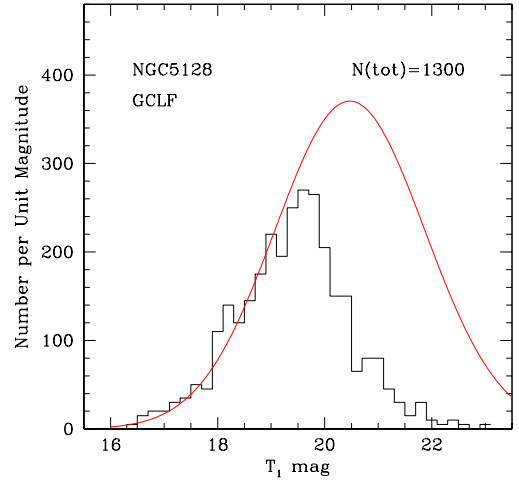}
\caption{Gaussian function superimposed on the 
NGC 5128 globular cluster luminosity function using data from 
\citet{wood10}. 
 }\label{fig.5}
\end{center}
\end{figure}

\subsection{GCS Metallicities and Ages}
\citet{har92} obtained Washington photometry to derive
metallicities for 62 GCs based on $C-T_1$ colours. The results
(Figure 6) 
showed a broad, bimodal metallicity distribution with  
a mean [Fe/H]$_{C-T_1}$ = -0.8$\pm$0.2.  None of the clusters showed 
the extreme, greater than solar abundances found by \citet{fro84}
based on $JHK$ photometry and it appears likely that these
values were due to the difficulty of calibration 
of colour-metallicity relations at such high abundances.


As the number of confirmed GCs has grown the ``simple'' bimodality
 of the MDF has become less clear (e.g. \citet{har04a}, \citet{peng04b},
\citet{bea08}, \citet{wood10} and references therein).  But the breadth
of the GCS colour distribution and its high mean metallicity
have not changed.  In addition, we now have chemical abundances and 
ages based on high S/N spectra which are helping to disentangle 
the frustrating age/metallicity degeneracy endemic to purely
photometric studies.  \citet{bea08} and \citet{wood10} have found 
that the majority of clusters, both metal-rich and
metal-poor, are old ($>$ 8-10Gyr).  The GCS has a subpopulation
of younger clusters ($<$3-5Gyr) but virtually no extremely young
members.  The clusters are $\alpha$-enhanced above solar, but not
as strongly as seen for Milky Way globulars. 

\subsection{Structure of GCs}
The structures of Milky Way globular clusters are well described 
by single mass, isotropic King models \citep{king62} and occupy
a narrow range of parameter space similar to the fundamental plane
for elliptical galaxies \citep{djor95}.  \citet{dean05} has
shown that the clusters have a common core M/L and that 
there is a strong correlation between core binding energy
and cluster luminosity.
\citet{barm07} and \citet{dean08} have shown that the GCs in
both M31 and NGC 5128 follow the same fundamental plane for clusters 
as massive as $\sim 3 \times 10^6 M_{\odot}$, suggesting
that old GCs have similar structural properties in all
galaxies.  Studies of the most luminous GCs in other systems 
\citep{has05,rej07} suggest that they bridge the gap between 
lower mass (normal) clusters and compact dwarfs (UCDs) \citep{mie08}. 
With a total population $\sim 10 \times$ that in the Milky Way, the 
NGC 5128 GCS is an important resource for exploring how structures
and M/L ratios change for the most massive clusters. 

\section{Field Halo Stars}
Our knowledge of the haloes of distant elliptical galaxies relies 
almost totally on measurements of their integrated light
(cf. the classic study of NGC 3379 by \citet{devau79}).
But the exceptional angular resolution 
and sensitivity of the HST changed
all that.  The first two-colour photometric study of an inner 
halo field in NGC 5128 \citep{sor96} used images that were
not very deep but did show the red giant tip and a broad
colour distribution for the giant stars.  The large angular 
size of NGC 5128 helps us in this kind of work, and we now have
deeper photometry for four fields at distances of $\sim$10, 20, 30, 
and 40 kpc or $\sim$1.5 - 7$r_{eff}$ (\citet{har99}, \citet{har00},
\citet{har02}, \citet{rej05}).

\begin{figure}[h]
\begin{center}
\includegraphics[scale=0.30, angle=0]{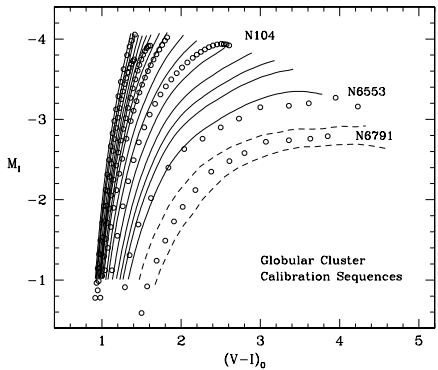}
\caption{Fiducial giant branches for Milky Way globular clusters
ranging in [m/H] from -2.0 to $\sim$ solar \citep{har02}.
 }\label{fig.6}
\end{center}
\end{figure}

\subsection{Metallicity of Halo Stars}
The brightest stars in an old stellar population are the 
red giants which first appear at $M_I \sim$-4.0.  Figure 6, 
 taken from \citet{har02},
shows this in the form of fiducial red giant branches (RGBs) 
for Milky Way GCs covering a metallicity range of 
$-2 \leq [m/H] \leq 0$.  From stellar model calculations
(e.g. \citet{VdB00}) we know the colour of the giant 
branch varies much less with age than metallicity.  
Therefore interpolating with RGB tracks calibrated onto the 
Milky Way giant branch (GB) grid provides a fast, efficient way to derive a 
first order metallicity distribution function (MDF).  We use $V,I$
photometry because it has a strong metallicity sensitivity 
for old stars; though not the best colour option, it 
represents a compromise between the best cameras available
and the spectral energy distributions of these old stars.

\begin{figure}[h]
\begin{center}
\includegraphics[scale=0.30, angle=0]{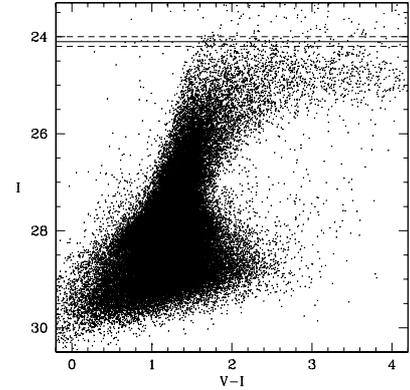}
\caption{Color-magnitude diagram for the outer halo of NGC 5128 
\citep{rej05},
dashed lines showing the TRGB measurement uncertainty of
 $\pm 0.1$ mag; compare with GC giant branches
plotted in Figure 6. }\label{fig.7}
\end{center}
\end{figure}

The $I,V-I$ colour magnitude diagram (CMD) for the outermost 
($\sim$40 kpc) NGC 5128 field \citep{rej05} is shown in Figure 7.  
This is the deepest of the halo star datasets but shows much the 
same GB characteristics as do the other three: a well defined 
upper boundary, especially on the blue (metal-poor) side of the 
giant branch and a very wide range in colour.  In Figure 8
are shown the MDFs for all four fields \citep{rej05}, with the data for
 the $\sim$ 20 and 30 kpc fields combined since they are essentially 
indistinguishable \citep{har02a}.  The RGB stars in NGC 5128's halo
are predominantly metal-rich with a broad MDF that varies little
with galactocentric distance.  And, in all fields, there are almost 
no metal-poor stars! 

\begin{figure}[h]
\begin{center}
\includegraphics[scale=0.30, angle=0]{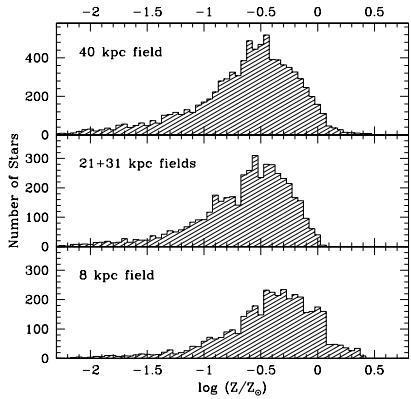}
\caption{Metallicity distributions for halo fields in NGC 5128 
\citep{rej05}.}
\label{fig.8}
\end{center}
\end{figure}

The lack of metal-poor stars out to almost 7$r_{eff}$ was a surprise
and naturally leads to the question: where is the
classic metal-poor halo in
this galaxy?  A possible clue to this puzzle can be seen in the $V,I$
 CMD of a field centred $\sim$ 30kpc or $\sim 12r_{eff}$
from the centre of the Leo elliptical NGC 3379.  At a distance
of $\sim$10 Mpc the HST ACS field covers a 
wide range in galactocentric distance of ~10kpc or 
$\sim 3 r_{eff}$.  Unlike what has been found in NGC 5128, 
the CMD for this field shows a significant 
metal-poor population along with the metal-rich stars that were 
expected \citep{har07b}.  When the image was divided into inner 
(closer to the galaxy centre) and outer (more distant), the CMDs
for these two regions were distinctly different.  As seen 
in Figure 9, the inner field shows a wide RGB colour range
which almost disappears in the outer field CMD.   In Figure 10 are 
plotted the radial distributions for the blue ([m/H]$<$-0.7) and red
([m/H]$>$-0.7] stars superimposed on the surface brightness
data from \citet{devau79}; the blue population density falls off with
radius as $\sigma \sim R^{-1.2}$ and the red as $\sigma \sim R^{-6.0}$.

\begin{figure}[h]
\begin{center}
\includegraphics[scale=0.30, angle=0]{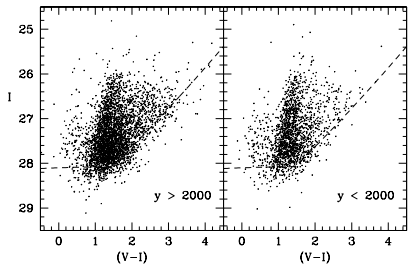}
\caption{CMD for a halo field in the Leo elliptical NGC 3379 \citep{har07b}.}
\label{fig.9}
\end{center}
\end{figure}

\begin{figure}[h]
\begin{center}
\includegraphics[scale=0.30, angle=0]{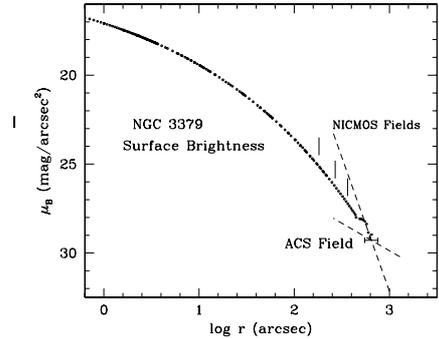}
\caption{Radial gradients for the blue and red field stars
in NGC 3379
from  \citet{har07b} superimposed on its surface brightness profile from
\citet{devau79}.}
\label{fig.10}
\end{center}
\end{figure}

Are we seeing here the transition to a classic metal-poor halo and, 
if so, why haven't we seen the same in NGC 5128?  Possibly the
reason is simply that the NGC 5128 data extend to only $\sim 7 r_{eff}$
compared with the NGC 3379 field which covered a range of
 $\sim 10.3-13.6 r_{eff}$.  Should we then expect the transition from 
a metal-rich to a metal-poor halo to occur at $\sim 10 r_{eff}$?  
Observational data which might help answer this question are limited, 
but a study by \citet{kal06} appears to have found such a 
transition for M31.  Their $V,I$ photometry for confirmed M31 halo
stars shows a radial gradient in metallicity, with the metal-poor 
component dominating beyond $r > 10r_{eff}$.
$V,I$ photometry for a halo field in NGC 5128 beyond $\sim 12 r_{eff}$ would 
be a simple way to test this.

\subsection{Age of Halo Stars}
The MDFs for the four halo fields described in the previous section
assumed a constant and old ($\sim$12 Gyr) age.  Elements of the CMDs 
such as the small numbers of stars brighter than the RGB tip and the 
lack of luminous blue stars suggest that this is a legitimate 
starting assumption.  But we would like to have better age constraints 
for these data, and this was the primary motivation for the 40kpc 
study (cf. Figure 7).  A CMD that reaches 
the main sequence turnoff is out of the
question at present, but it was possible to obtain data to the depth 
of the horizontal branch ($I = 28-29$).  The resulting photometry
is good enough to compare with CMD simulations 
and narrow the possible range in age and metallicity using red clump and
asymptotic giant branch stars as well as those on the RGB.  It
now appears that the dominant population in this outer field has an 
age of $\sim$ 10-12Gyr and is probably combined with a small
fraction of younger ($\sim$ 5Gyr) stars.

\begin{figure}[h]
\begin{center}
\includegraphics[scale=0.30, angle=0]{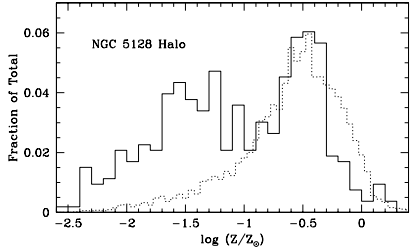}
\caption{Comparison of GC (solid histogram) and field star (dotted
 histogram) MDFs in NGC 5128 from \citet{harris09}.}
\label{fig11}
\end{center}
\end{figure}

A comparison between the GC and halo field stars (cf.
Figure 11 of \citet{harris09}) shows some intriguing 
similarities.  The metallicity peaks for the halo stars and the metal-rich 
GCSs are quite similar and current analysis indicates that both samples 
are mainly old with a small younger component.  Simplistically this
suggests that major star formation episodes at $\sim$ 12 and 5Gyr 
may have occurred in NGC 5128.

\section{Summary}
$\bullet$ There is now agreement, within a few $\%$ as to the distance 
to NGC 5128, making it possible to interpret observed 
properties with new confidence. \\
$\bullet$ Although there is still debate
as to whether this is an S0p or Ep galaxy, the properties 
of its halo and individual stars are consistent with classifying
NGC 5128 as an Ep. \\
$\bullet$ We have a sample of $\sim$600 globular clusters
confirmed by radial velocity and/or angular resolution, approximately
50$\%$ of the estimated total population.  Roughly half of these are 
metal-rich and the majority are old.  \\
$\bullet$ In spite of these numbers,
global properties of the system are not well known because of
major biases in the spatial coverage. \\
$\bullet$ Observations of the spatial structure  
of $>$200 GCs shows that they follow the same fundamental plane
relation as do GCs in the Milky Way and M31. The NGC 5128 system allows
us to trace this relation to higher GC luminosity and mass than 
we can in the Milky Way or M31. \\
$\bullet$ Over a wide range in
galactocentric distance, the halo stars are remarkably similar and 
predominantly metal-rich; and it appears that these metal-rich stars
are old, possibly with ages similar to those of the metal-rich GCs.  
We may need to observe the halo to galactocentric 
distances twice that of current 
datasets in order to uncover its metal-poor component.



\section*{Acknowledgments} 
GLHH acknowledges financial support through a research grant from the
Natural Sciences and Engineering Research Council of Canada.   The 
beginnings of this paper were developed during a visit to ESO
Garching, sponsored through the ESO visitor programme.


\end{document}